\documentclass[preprint,11pt,showpacs,amsmath,amssymb,floatfix]{revtex4}

\usepackage{amsbsy}
\usepackage{amssymb}
\usepackage{color}
\usepackage{epsfig}
\usepackage{hyperref} 
\usepackage{graphicx}
\usepackage{subfigure}
\usepackage{soul}

\newcommand{\beq}{\begin{equation}}

\newcommand{\eeq}{\end{equation}}

\begin{document}

\title{Massive photons from Super and Lorentz symmetry breaking}

\author{Luca Bonetti\textsuperscript{1,2}, Lu{\'i}s R. dos Santos Filho\textsuperscript{3}, Jos{\'e} A. Helay{\"e}l-Neto\textsuperscript{3}, Alessandro D.A.M. Spallicci\textsuperscript{1,2}\footnote{Corresponding author: spallicci@cnrs-orleans.fr}
}
\affiliation{
\textsuperscript{1}Universit\'e d'Orl\'eans\\
\mbox{Observatoire des Sciences de l'Univers en r\'egion Centre (OSUC) UMS 3116} \\
\mbox{1A rue de la F\'{e}rollerie, 45071 Orl\'{e}ans, France}\\
\mbox{Collegium Sciences et Techniques (COST), P\^ole de Physique}\\
\mbox{Rue de Chartres, 45100  Orl\'{e}ans, France}\\
\textsuperscript{2}Centre Nationale de la Recherche Scientifique (CNRS)\\
\mbox{Laboratoire de Physique et Chimie de l'Environnement et de l'Espace (LPC2E) UMR 7328}\\
\mbox {Campus CNRS, 3A Av. de la Recherche Scientifique, 45071 Orl\'eans, France}
\vskip3pt
\textsuperscript{3}Centro Brasileiro de Pesquisas F\'{\i}sicas (CBPF)\\
\mbox {Rua Xavier Sigaud 150, 22290-180 Urca, Rio de Janeiro, RJ, Brasil}}

\date{29 July 2016}

\begin{abstract}
In the context of Standard Model Extensions (SMEs), we analyse four general classes of 
Super Symmetry (SuSy) and Lorentz Symmetry (LoSy) breaking, leading to {observable} imprints at our energy scales. The photon dispersion relations show a non-Maxwellian behaviour for the CPT (Charge-Parity-Time reversal symmetry) odd and even sectors. The group velocities exhibit also a directional dependence with respect to the breaking background vector (odd CPT) or tensor (even CPT).  In the former sector, the group velocity may decay following an inverse squared frequency behaviour. Thus, we extract a massive and gauge invariant Carroll-Field-Jackiw photon term in the Lagrangian and show that the mass is proportional to the breaking vector. The latter is estimated by ground measurements and leads to a photon mass upper limit of $10^{-19}$ eV or $2 \times 10^{-55}$ kg and thereby to a potentially measurable delay at low radio frequencies. 
\end{abstract}

\maketitle

We largely base our understanding of particle physics on the Standard Model (SM). Despite having proven
to be a very reliable reference, there are still unsolved problems, {such as} the Higgs Boson 
 mass overestimate, the absence of a candidate particle for the dark universe, {as well as} the neutrino oscillations and their mass.


Standard Model {Extensions (SMEs)} tackle these problems. Among {them}, SuperSymmetry (SuSy) \cite{fayet2014,terning2006} figures new physics {at} TeV scales \cite{lykken2010}. Since, in SuSy{,} Bosonic and Fermionic particles {each have} a counterpart, their mass {contributions} cancel each other and allow the correct experimental low mass value for the Higgs Boson.

Lorentz Symmetry (LoSy) is assumed in the SM. It emerges  \cite{kosteleckysamuel1989a,kosteleckysamuel1991,kosteleckypotting1991,kosteleckypotting1996}
that in the context of Bosonic strings, the condensation of tensor fields
is dynamically possible and determines LoSy violation. 
There are opportunities to test the low energy manifestations of LoSy violation, through SMEs \cite{colladaykostelecky1997,colladaykostelecky1998}. The effective Lagrangian is given by the usual SM Lagrangian corrected by SM operators of any dimensionality contracted with suitable Lorentz breaking tensorial (or simply vectorial) background coefficients. In this letter, we show that photons exhibit a non-Maxwellian behaviour, are massive and possibly manifest dispersion at low frequencies, pursued by newly operating ground radio observatories and future space missions.    

LoSy violation has been analysed phenomenologically. Studies include electrons, photons, muons, mesons,
baryons, neutrinos and Higgs sectors. Limits on the parameters associated {with} the breaking of relativistic
covariance are set by {numerous} {analyses} \cite{baileykostelecky2006,kosteleckytasson2009,kosteleckyrussell2011}, including {with} electromagnetic cavities and optical systems \cite{bestwakola2000,phhumastvewa2001,huphmavest2003,mubrhepela2003,muhesapela2003,muller2005,russell2005}.
Also Fermionic strings have been proposed in {the} presence of LoSy violation. Indeed, the magnetic properties of spinless and/or neutral particles with a non-minimal coupling to a LoSy violation background have been {placed} in relation to Fermionic matter or gauge Bosons 
\cite{casanaferreirasantos2008a,casanaferreirarodriguessilva2009, bakkebelichsilva2011a,besifeor2011,limabelichbakke2013,hernaskibelich2014}.

\begin{figure}[h!]
\centering \includegraphics[width=16cm]{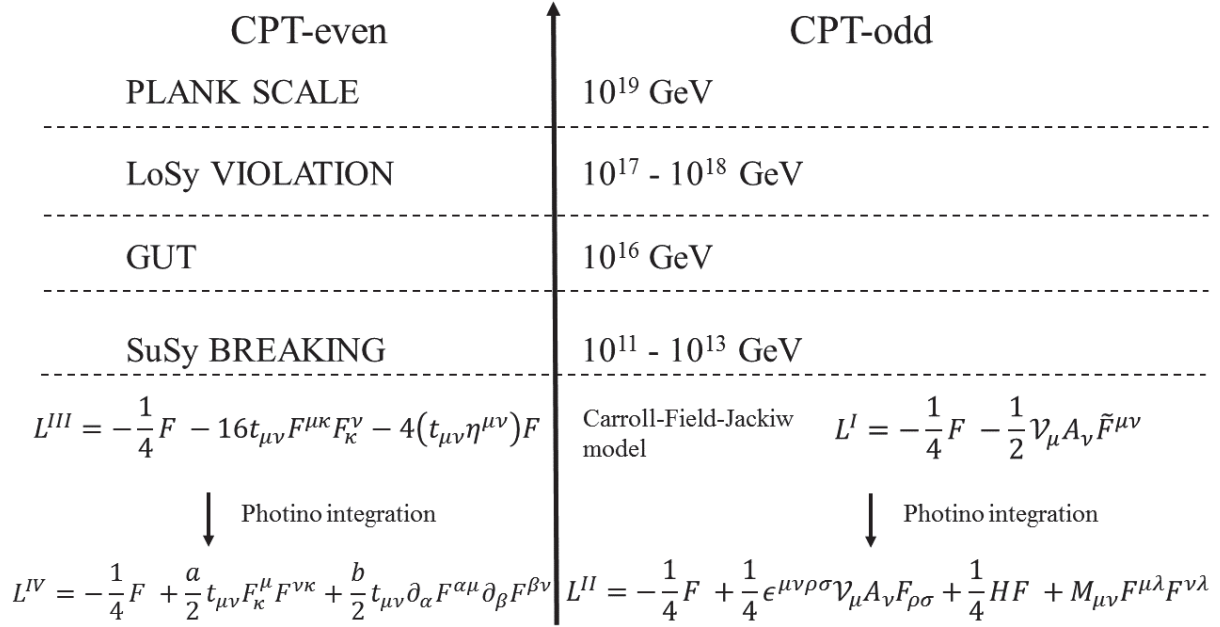}
\caption{Breaking energy values and the Lagrangians. A different hierarchy of LoSy, SuSy breaking and Grand Unification Theories (GUT) does not interfere with the dispersion laws of the photonic sector at low energies.}
\label{ev}
\end{figure}

LoSy violation occurs at larger energy scales than those obtainable in particle accelerators \cite{bergerkostelecky2002,nibbelinkpospelov2005,katzshadmi2006,hnbedilesp2010,falenape2012,redigolo2012,golenapeds2013}. At those energies, SuSy is still an exact symmetry, even if we assume that it might break at scales close to the primordial ones. However, LoSy violation naturally induces SuSy breaking because the background vector (or tensor) - that implies the LoSy violation - is in fact part of a SuSy multiplet \cite{bebegahnle2015}, Fig. (\ref{ev}).

The sequence is assured by the supersymmetrisation, in the  CPT (Charge-Parity-Time reversal symmetry) odd sector, of the  Carrol-Field-Jackiw (CFJ) model \cite{cafija90} that emulates a Chern-Simons \cite{dunne1998} term and includes a background field that breaks LoSy, under the point of view of the so-called (active)
particle transformations. The latter consists of transforming the potential $A^\mu$ and the field $F^{\mu\nu}$, while keeping the background vector ${\cal V}^\mu$ unchanged. For the photon sector, when unaffected by the photino contribution, the CFJ Lagrangian reads (Class I)

\begin{align}
&
L^{I}=-\frac{1}{4}F-\frac{1}{2}{\cal V}_{\mu}A_{\nu}\tilde{F}^{\mu\nu}~,
\label{lagrangian1}
\\
&
\tilde{F}^{\mu\nu}=\frac{1}{2}\epsilon^{\mu\nu\alpha\beta}F_{\alpha\beta}~,
\label{effemunu}
\end{align}
where $F = F^{\mu\nu}F_{\mu\nu}$. The term in Eq. (\ref{effemunu}) couples the photon to an external constant four vector and it violates parity even if gauge
symmetry is respected \cite{cafija90}. If the CFJ model is supersymmetrised \cite{bebegahn2013}, the vector ${\cal V}^{\mu}$ is space-like constant and is given by the gradient of the SuSy breaking scalar background field, present in the matter supermultiplet. The dispersion relation yields, {denoting} 
$k^{\mu}=(\omega,\vec{k})$, $k^2 = (\omega^{2}-|\vec{k}|^{2})$, and $({\cal V}^{\mu}k_{\mu})^{2}= 
({\cal V}^{0}\omega - \vec{\cal V}\cdot\vec{k})^2${,}


\beq
k^{4}+{\cal V}^{2}k^{2}-({\cal V}^{\mu}k_{\mu})^{2}=0~. 
\label{disp1}
\eeq

If SuSy holds and the photino degrees of freedom are integrated out, we are led to the effective photonic action, {\it i.e.} the effect of the photino on the photon propagation. The Lagrangian (\ref{lagrangian1}) is recast as (Class II) \cite{bebegahnle2015} 

\begin{equation}
L^{II}=-\frac{1}{4}F+\frac{1}{4}\epsilon^{\mu\nu\rho\sigma}{\cal V}_{\mu}A_{\nu}F_{\rho\sigma}+\frac{1}{4}HF+M_{\mu\nu}F^{\mu\lambda}F^{\nu\lambda}~,
\label{Lagrangian2}
\end{equation}
where $H$, the tensor $M_{\mu\nu}=\tilde{M}_{\mu\nu}+1/4 \eta_{\mu\nu}M$, and $\tilde{M}_{\mu\nu}$ depend on the background Fermionic
condensate, originated by SuSy; $M_{\mu\nu}$ is traceless, $M$ is the trace of $M_{\mu\nu}$ and $\eta_{\mu\nu}$ the metric. Thus, the Lagrangian, Eq. (\ref{Lagrangian2}), in terms of the irreducible terms displays as
\beq
L^{II}=-\frac{1}{4}\left(1-H-M\right)F+\frac{1}{4}\epsilon^{\mu\nu\rho\sigma}{\cal V}_{\mu}A_{\nu}F_{\rho\sigma}+\tilde{M}_{\mu\nu}F^{\mu\lambda}F_{\lambda}^{\nu}~.
\eeq

The corresponding dispersion relation reads
\beq
k^4 +\frac{{\cal V}^{2}}{(1-H-M)^{2}}k^2  
 -\frac{1}{(1-H-M)^{2}}{\cal V}^{\mu}k_{\mu}=0~. 
\label{disp2}
\eeq


The dispersion law {given} by Eq. (\ref{disp2}) is just a rescaling of Eq. (\ref{disp1}) as we integrated out the photino sector. The background parameters are very small{,} being suppressed exponentially at the Planck scale{; they} render the denominator in 
Eq. (\ref{disp2}) close to unity, implying similar numerical outcomes for the two {dispersions} of Classes I and II.

The even sector \cite{bebegahnle2015} assumes that the Bosonic background, responsible of LoSy violation, is a background tensor $t_{\mu\nu}$. For the photon sector, if unaffected by the photino contribution, the Lagrangian reads (Class III)

\beq
L^{III}=-\frac{1}{4}F-16t_{\mu\nu}F^{\mu\kappa}F_{\kappa}^{\nu}-4\left(t_{\mu\nu}\eta^{\mu\nu}\right)F~.
\label{Lagrangian3}
\eeq

The dispersion relation for Class III \cite{bodshnsp2016} is 

\beq
\omega^{2}-\left(1+\rho+\sigma\right)^{2}|\vec{k}|^{2}=0~,
\label{Dispersion3}
\eeq
where $ \rho  = 1/2 \tilde{K}_{\alpha}^{\alpha}$, $\sigma  =  1/2\tilde{K}^{\alpha\beta}\tilde{K}_{\alpha\beta}-\rho^{2}$, and 
$\tilde{K}^{\alpha\beta}  =  t^{\alpha\beta}t^{\mu\nu}{p_{\mu}p_{\nu}}/|\vec{k}|^{2}${ are} associated to Fermionic condensates.

Integrating out the photino \cite{bebegahnle2015}, we turn to the Lagrangian of Class IV

\begin{equation}
L^{IV}=-\frac{1}{4}F + \frac{a}{2}t_{\mu\nu}F_{\kappa}^{\mu}F^{\nu\kappa}+\frac{b}{2}t_{\mu\nu}\partial_{\alpha}F^{\alpha\mu}\partial_{\beta}F^{\beta\nu}~,
\label{Lagrangian4}
\end{equation}
{where} $a$ {is }a dimensionless coefficient and $b$ a parameter of dimension of mass$^{-2}$ (herein, $ c=1$, unless otherwise stated). 
For the dispersion relation, we write the Euler-Lagrange equations,
pass {to Fourier space} and set to zero the determinant of the matrix that multiplies the Fourier transformed potential.
However, given the complexity of the matrix in this case  and the smallness of the tensor $t_{\mu\nu}$, we develop the determinant in {a} series truncated at first order and get \cite{bodshnsp2016} 

%
%

\beq
btk^{4}-k^{2}+\left(3a+bk^{2}\right)t^{\alpha\beta}k_{\alpha}k_{\beta}=0~,
\label{Dispersion4}
\eeq
where $t=t_{\mu}^{\mu}$.

For determining the group velocity, we first consider ${\cal V}_{0}=0$ for Class I \cite{adamklinkhamer2001,bsbebohn2003} and obtain

\beq
\omega^{4}-\left(2|\vec{k}|^{2}+|\vec{\cal V}|^2\right)\omega^{2}+|\vec{k}|^{4}+|\vec{k}|^{2}|\vec{\cal V}^2-\left(\vec{\cal V}\cdot\vec{k}\right)^{2}=0~.
\label{omega4vgIV00}
\eeq

In \cite{bsbebohn2003}, the authors do not exploit the consequences of the dispersion relation and do not consider a SuSy scenario. 
Dealing with Eq. (\ref{omega4vgIV00}), we have neglected the negative roots; it turns out that the two positive roots determine identical group velocities $dw/dk$ up to second order in $\vec{\cal V}$.
For $\theta$, {the} angle between the background vector $\vec{\cal V}$ and $\vec{k}$, we get 
\beq
v_{g}^{I}|^{\theta \neq \pi/2}_{{\cal V}_0=0} = 
1 - \frac{|\vec{\cal V}|^2}{8\omega^2} (2 + \cos^2\theta)~,
\label{vgIV00anytheta}
\eeq
for $\theta \neq \pi/2$. Instead for $\theta = \pi/2$, one of the two solutions coincides with the Maxwellian value, while the other is dispersive 

\beq
v_{g}^{I1}|^{\theta = \pi/2}_{{\cal V}_0=0} = 1~,
~~~~~~~~~~~~~~~ 
v_{g}^{I2}|^{\theta = \pi/2}_{{\cal V}_0=0} = 1 - \frac{1}{2}\frac{|\vec{\cal V}|^2}{\omega^2}~.
\label{vgIV00thetapiover2}
\eeq

 
For ${\cal V}_{0}\neq 0$, we suppose that the light propagates along the z axis ($k_1 = k_2 = 0$) which for convenience is along the line of sight of the source. We then obtain 

\beq
\omega^4 - [2k_3^2+{\cal V}_1^2 + {\cal V}_2^2 + {\cal V}_3^2]\omega^2 + 2 {\cal V}_0 {\cal V}_3 k_3~\omega 
 + k_3^4 + ({\cal V}_1^2 + {\cal V}_2^2 - {\cal V}_0^2) k_{3}^{2} =0~.
\label{omega4}
\eeq 


We now set ${\cal V}_3=0$, that is{, the light propagates} orthogonally to the background vector.  
Further, {for ${\cal V}$ spacelike and} 
$4 {\cal V}_0^2k_3^2/|\vec{\cal V}|^4 \ll 1$, we get two group velocities, one of which is dispersive 



\beq
v_g^{I1}|_{{\cal V}_3=0} = 1 - \frac{{\cal V}_0^2}{|\vec{\cal V}|^2}~, 
~~
v_g^{I2}|_{{\cal V}_3=0}  
\simeq
 \alpha  \left( 1 - \frac{1}{2}\frac{|\vec{\cal V}|^2}{\omega^2}\right)~.
\label{vgI1V30}
\eeq

The solution $v_g^{I1}|_{{\cal V}_3=0}$ is always subluminal {for} ${\cal V}$ spacelike. 
The solution $v_g^{I2}|_{{\cal V}_3=0}$ assumes $\omega \gg |\vec{\cal V}|$. Since $\alpha = 1 + {\cal V}_0^2/|\vec{\cal V}|^2$,  
$v_g^{I2}|_{{\cal V}_3=0}$ is superluminal for $ \sqrt{2} \omega > |\vec{\cal V}| (1+ |\vec{\cal V}|^2/{\cal V}^2_0)^{1/2} $. Further, the value of $\alpha$ is not Lorentz-Poincar\'e invariant. Superluminal behaviour is avoided assuming for both solutions ${\cal V}_0=0$.

\begin{figure}[h!]
\centering
\includegraphics[width=16cm]{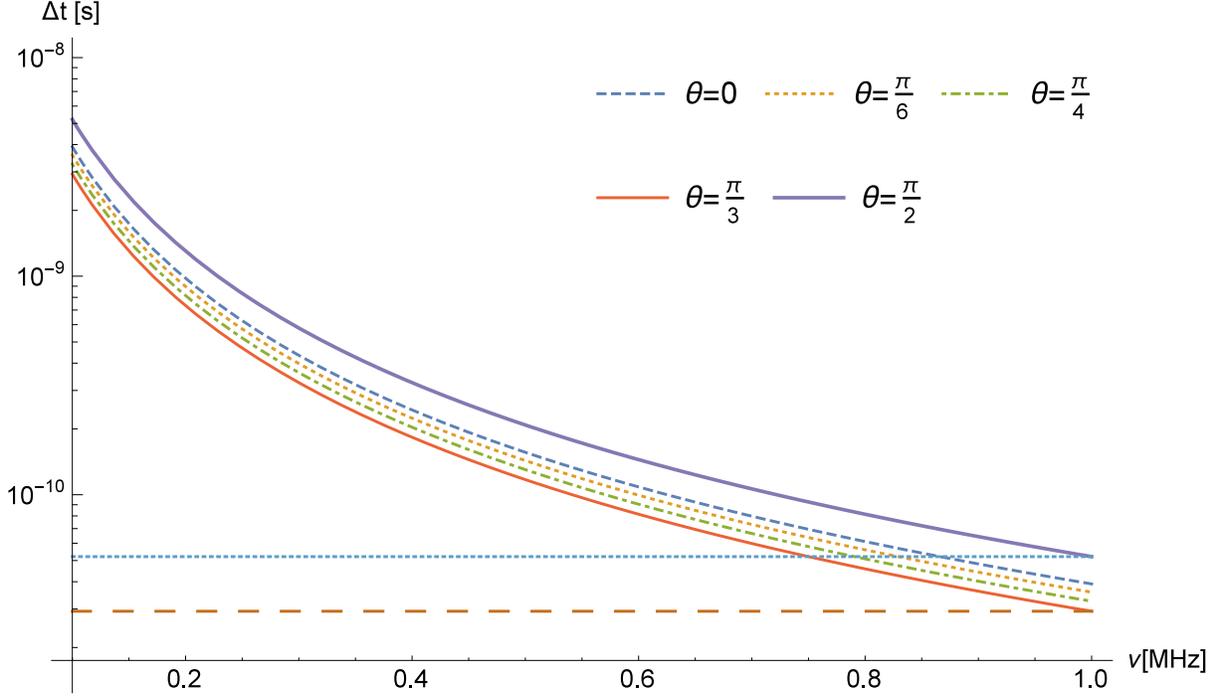}
\caption{For Class I, we plot the delays [s], Eq. (\ref{deltatCFJ}), for different angles, Eqs. (\ref{vgIV00anytheta},\ref{vgIV00thetapiover2}), {using} $|\vec{\cal V}| = 10^{-19}$ eV \cite{gomesmalta2016}, versus frequency. We have supposed the source {to be} at a distance of 4 kpc.
The frequency range 0.1 - 1 MHz has been chosen since {it is} targeted by recently proposed low radio frequency space detectors, composed by a swarm of nano-satellites{;} see \cite{bebosp2016} and references therein. There is a feeble dependence of the delays on $\theta$. 
The delay is of about 50 ps at 1 MHz for $\theta = \pi/2$, Eq. (\ref{vgIV00thetapiover2}), and around {half} of this value for 
$\theta$ approaching $\pi/2$, Eq. (\ref{vgIV00anytheta}).}
\label{delay}
\end{figure}

If dealing only with a null ${\cal V}_0$ and with dispersive group velocities, for a source at distance $\ell$, the time delay of two photons at different frequencies, A and B, is given by (in SI units) 

\beq
\Delta t_{\rm CFJ} = \frac{\ell |\vec{\cal V}|^2}{2 c \hbar^2}\left( \frac{1}{\omega^2_A} - \frac{1}{\omega^2_B}\right) x~,
\label{deltatCFJ}
\eeq
where $x$ takes the values $(2 + \cos^2\theta)/4$, for Eq. (\ref{vgIV00anytheta}), and 1 for Eqs. (\ref{vgIV00thetapiover2},\ref{vgI1V30}). 
The delays, Eq. (\ref{deltatCFJ}), are plotted in Fig. (\ref{delay}).    
Comparing with the de Broglie-Proca (dBP) delay 

\beq
\Delta t_{\rm dBP} = \frac{\ell m_\gamma^2 c^3}{2 \hbar^2}\left( \frac{1}{\omega^2_A} - \frac{1}{\omega^2_B}\right)~,
\label{deltatdBP}
\eeq
we conclude that the background vector induces an effective mass {to the photon, $m_\gamma$, of value}

\beq
m_\gamma = \frac{|\vec{\cal V}|}{c^2}x~.
\label{mgamma}
\eeq

Equation (\ref{mgamma}) is gauge-invariant, conversely to the potential dependent dBP mass. It appears {as} the pole of the transverse component of the photon propagator \cite{bsbebohn2003}.



Class II, just a rescaling of Class I, implies identical solutions, differing {by} a numerical factor only.

%


The group velocities of Classes III and IV show no sign of dispersion; they are slightly smaller than c - as light travelling through matter, but suffer {from} anisotropy to a larger degree than in Classes I and II. Indeed,  the isotropy is lost due to the tensorial nature of the LoSY and SuSy breaking perturbation. The feebleness of the corrections is due to the coefficient ${\cal{T}}$ being proportional to the powers of the tensor $t_{\mu\nu}$ components, of $10^{-19}$ eV magnitude \cite{bodshnsp2016}  

\beq
v_g^{III,IV} = 1 - {\cal{T}}\left(t_{1}\mbox{sin}^{2}\theta\mbox{cos}^{2}\varphi+t_{2}\mbox{sin}^{2}\theta\mbox{sin}^{2}\varphi+t_{3}\mbox{cos}^{2}\theta\right)~,
\eeq
where $\theta$ and $\varphi$ are the azimuthal and planar angles of $\vec{k}$ with respect to the axes respectively.

Having seen a typical dBP massive photon behaviour in the group velocities of the odd sector, we rewrite Eq. (\ref{lagrangian1}) in terms of the potentials {to let a} massive CFJ photon {\it \`a la} de Broglie-Proca {emerge}

\beq
L=\frac{1}{2}\left(\vec{\nabla}\phi+\dot{\vec{A}}\right)^{2}-\frac{1}{2}\left(\vec{\nabla}\times\vec{A}\right)^{2}+{\cal V}_{0}\vec{A}\cdot\left(\vec{\nabla}\times\vec{A}\right)
-2\vec{\nabla}\phi\cdot\left(\vec{\cal V}\times\vec{A}\right)-\vec{\cal V}\cdot\left(\vec{A}\times\dot{\vec{A}}\right)~.\nonumber
\eeq


Since the $\phi$ field appears only through its gradient, {in} the absence of $\phi$ time derivatives and thereby of dynamics, 
 $\vec{\nabla}\phi$ acts as an auxiliary field and {can} be integrated out from the Lagrangian. Defining 
$\chi=\vec{\nabla}\phi+\dot{\vec{A}}-2\vec{\cal V}\times\vec{A}$, we get

\beq
L=\frac{1}{2}\chi^{2}-2\left(\vec{\cal V}\times\vec{A}\right)^{2}+\vec{\cal V}\cdot\left(\vec{A}\times\dot{\vec{A}}\right)
-\frac{1}{2}\left(\vec{\nabla}\times\vec{A}\right)^{2}+{\cal V}_{0}\vec{A}\cdot\left(\vec{\nabla}\times\vec{A}\right)~.
\label{CFJpots}
\eeq


The Euler-Lagrangian equation for $\chi$ is disregarded since $\chi=0$.  The term $\left(\vec{\cal V}\times\vec{A}\right)^{2} $ is expanded as $\left(\left|\vec{\cal V}\right|^{2}\delta_{kn}-{\cal V}_{k}{\cal V}_{n}\right)A_{k}A_{n}:=M_{kn}\left(\vec{\cal V}\right)A_{k}A_{n}$, where $M_{kn}$ is a symmetric diagonalisable matrix, thanks to a suitable matrix of the $SO\left(3\right)$ rotation group.
Performing such a change in Eq. (\ref{CFJpots}), the term under discussion changes into 
\beq
\tilde{A}_{i}\tilde{M}_{ij}\tilde{A_{j}=}\left|\vec{\cal V}\right|^{2}\tilde{A}_{2}^{2}+\left|\vec{\cal V}\right|^{2}\tilde{A}_{3}^{2}~,
\eeq
thereby showing a massive photon term like in the de Broglie-Proca Lagrangian. 

The quest for a photon with non vanishing mass is 
definitely not new. The first attempts can be traced back to de Broglie who conceived an upper limit of $10^{-53}$ kg, 
and achieved a comprehensive formulation of the photon \cite{db40}, also thanks to the reinterpretation of the work of his doctorate student Proca. To the Lagrangian of Maxwell's electromagnetism, they added a gauge breaking term proportional to the square of the photon mass. A laboratory Coulomb's law test determined the mass upper limit of $2\times 10^{-50}$~kg \cite{wifahi71}. In the solar wind, Ryutov found $10^{-52}$~kg at 1 AU \cite{ry97,ry07}, and $1.5\times 10^{-54}$~kg at $40$ AU \cite{ry07}. These  
limits were accepted by the Particle Data Group (PDG) \cite{oliveetal2014}, but recently put into 
question \cite{retinospalliccivaivads2016}\footnote{In \cite{retinospalliccivaivads2016}, there are references to even more optimistic, and less safe, limits ($3 \times 10^{-63}$ kg), claimed when modelling the galactic magnetic field. For recent studies on propagation limits see \cite{boelmasasgsp2016,wuetal2016b}.}.
The lowest value for any mass is dictated by Heisenberg's principle $m \geq \hbar/\Delta t c^2$, and gives $1.3\times 10^{-69}$~kg,
where $\Delta t$ is the supposed age of the Universe.

In this letter, we have focused on Susy and LoSy breaking and derived the ensuing dispersion relations and group velocities for four types of {Lagrangians}. All group velocities show a non-Maxwellian behaviour, in the angular dependence and through sub or super luminal speeds.  
Superluminal behaviour is exclusive to the odd CPT sector, and may occur only if the time component 
of the perturbing vector is non-null. Further, in the odd CPT sector, the mass 
shows a dispersion, proportional to $1/\omega^2$, as in dBP formalism. The difference lies in the gauge invariance of the CFJ photon, the mass of which is proportional to $|\vec{\cal V}|$. 
The delays are thus more important at lower frequencies and the opening of the 0.1-100 MHz window would be of importance \cite{bebosp2016}. 
Elsewhere, we have analysed the polarisation and evinced the transversal and longitudinal (massive) modes \cite{bodshnsp2016}.  

From the rotation of the plane of polarisation from distant galaxies, or from the Cosmic Microwave Background (CMB), it has been assessed {that} 
$|{\cal V}_\mu| < 10^{-34}$ eV \cite{cafija90,goldhabertrimble1996,kosteleckyrussell2011}. This result is comparable to the Heisenberg mass limit value at the age of the universe. 
A less stringent, but interesting, limit of $10^{-19}$ eV \cite{gomesmalta2016} has been set through laboratory based experiments involving electric dipole moments of charged leptons or {the} inter-particle potential between Fermions and the associated corrections to the spectrum{ of} the Hydrogen atom. These latter estimates imply, Eq. (\ref{mgamma}), a mass upper limit of $10^{-55}$ kg. 

The detection of the CFJ massive photon can be pursued by other means, {\it e.g.}, through analysis of Amp\`ere's law
in the solar wind \cite{retinospalliccivaivads2016}. Incidentally, the odd and even CPT sectors can be experimentally separable
 \cite{kosteleckyrussell2011}.  

What is the role of a massive photon for SMEs? 
String theory has hinted to massive gravitons and photons \cite{kosteleckypotting1991,kosteleckysamuel1991}, while Proca electrodynamics was investigated in the context of LoSy violation, but outside a SuSy scenario \cite{casanaferreirasantos2008a}. However, if LoSy takes place in a supersymmetric scenario, the photon mass may be naturally generated from SuSy breaking condensates 
\cite{bebegahn2013,bebegahnle2015}.    
As a final comment, we point out that the emergence of a massive photon is pertinent also to other SME formulations. 

LB and ADAMS acknowledge CBPF for hospitality, while LRdSF and JAHN are grateful to CNPq-Brasil for financial support.

\bibliography{references_spallicci_160728}

\begin{thebibliography}{50}
\expandafter\ifx\csname natexlab\endcsname\relax\def\natexlab#1{#1}\fi
\expandafter\ifx\csname bibnamefont\endcsname\relax
  \def\bibnamefont#1{#1}\fi
\expandafter\ifx\csname bibfnamefont\endcsname\relax
  \def\bibfnamefont#1{#1}\fi
\expandafter\ifx\csname citenamefont\endcsname\relax
  \def\citenamefont#1{#1}\fi
\expandafter\ifx\csname url\endcsname\relax
  \def\url#1{\texttt{#1}}\fi
\expandafter\ifx\csname urlprefix\endcsname\relax\def\urlprefix{URL }\fi
\providecommand{\bibinfo}[2]{#2}
\providecommand{\eprint}[2][]{\url{#2}}

\bibitem[{\citenamefont{Fayet}(2014)}]{fayet2014}
\bibinfo{author}{\bibfnamefont{P.}~\bibnamefont{Fayet}}, \bibinfo{journal}{Eur.
  J. Phys. C} \textbf{\bibinfo{volume}{74}}, \bibinfo{pages}{2837}
  (\bibinfo{year}{2014}), \eprint{arXiv:hep-ph/0111282}.

\bibitem[{\citenamefont{Terning}(2006)}]{terning2006}
\bibinfo{author}{\bibfnamefont{J.}~\bibnamefont{Terning}},
  \emph{\bibinfo{title}{Modern supersymmetry - dynamics and duality}}
  (\bibinfo{publisher}{Oxford Science Publications}, \bibinfo{address}{Oxford},
  \bibinfo{year}{2006}).

\bibitem[{\citenamefont{Lykken}(2010)}]{lykken2010}
\bibinfo{author}{\bibfnamefont{J.~D.} \bibnamefont{Lykken}}, in
  \emph{\bibinfo{booktitle}{Proc. of the {E}uropean school of high energy
  physics}}, edited by
  \bibinfo{editor}{\bibfnamefont{C.}~\bibnamefont{Grojean}} \bibnamefont{and}
  \bibinfo{editor}{\bibfnamefont{M.}~\bibnamefont{Spiropulu}}
  (\bibinfo{publisher}{CERN}, \bibinfo{address}{Gene{\'{e}}ve},
  \bibinfo{year}{2010}), {Y}ellow {R}eport CERN-2010-002, p.
  \bibinfo{pages}{101}, \bibinfo{note}{arXiv:1005.1676 [hep-ph], 14-27 June
  2009 Bautzen}.

\bibitem[{\citenamefont{Kosteleck{\'{y}} and
  Samuel}(1989)}]{kosteleckysamuel1989a}
\bibinfo{author}{\bibfnamefont{V.~A.} \bibnamefont{Kosteleck{\'{y}}}}
  \bibnamefont{and} \bibinfo{author}{\bibfnamefont{S.}~\bibnamefont{Samuel}},
  \bibinfo{journal}{Phys. Rev. Lett.} \textbf{\bibinfo{volume}{63}},
  \bibinfo{pages}{224} (\bibinfo{year}{1989}).

\bibitem[{\citenamefont{Kosteleck{\'{y}} and
  Samuel}(1991)}]{kosteleckysamuel1991}
\bibinfo{author}{\bibfnamefont{V.~A.} \bibnamefont{Kosteleck{\'{y}}}}
  \bibnamefont{and} \bibinfo{author}{\bibfnamefont{S.}~\bibnamefont{Samuel}},
  \bibinfo{journal}{Phys. Rev. Lett.} \textbf{\bibinfo{volume}{66}},
  \bibinfo{pages}{1811} (\bibinfo{year}{1991}).

\bibitem[{\citenamefont{Kosteleck{\'{y}} and
  Potting}(1991)}]{kosteleckypotting1991}
\bibinfo{author}{\bibfnamefont{V.~A.} \bibnamefont{Kosteleck{\'{y}}}}
  \bibnamefont{and} \bibinfo{author}{\bibfnamefont{R.}~\bibnamefont{Potting}},
  \bibinfo{journal}{Nucl. Phys. B} \textbf{\bibinfo{volume}{359}},
  \bibinfo{pages}{545} (\bibinfo{year}{1991}).

\bibitem[{\citenamefont{Kosteleck{\'{y}} and
  Potting}(1996)}]{kosteleckypotting1996}
\bibinfo{author}{\bibfnamefont{V.~A.} \bibnamefont{Kosteleck{\'{y}}}}
  \bibnamefont{and} \bibinfo{author}{\bibfnamefont{R.}~\bibnamefont{Potting}},
  \bibinfo{journal}{Phys. Lett. B} \textbf{\bibinfo{volume}{381}},
  \bibinfo{pages}{89} (\bibinfo{year}{1996}), \eprint{arXiv:hep-th/9605088}.

\bibitem[{\citenamefont{Colladay and
  Kosteleck{\'{y}}}(1997)}]{colladaykostelecky1997}
\bibinfo{author}{\bibfnamefont{D.}~\bibnamefont{Colladay}} \bibnamefont{and}
  \bibinfo{author}{\bibfnamefont{V.~A.} \bibnamefont{Kosteleck{\'{y}}}},
  \bibinfo{journal}{Phys. Rev. D} \textbf{\bibinfo{volume}{55}},
  \bibinfo{pages}{6760} (\bibinfo{year}{1997}), \eprint{arXiv:hep-ph/9703464}.

\bibitem[{\citenamefont{Colladay and
  Kosteleck{\'{y}}}(1998)}]{colladaykostelecky1998}
\bibinfo{author}{\bibfnamefont{D.}~\bibnamefont{Colladay}} \bibnamefont{and}
  \bibinfo{author}{\bibfnamefont{V.~A.} \bibnamefont{Kosteleck{\'{y}}}},
  \bibinfo{journal}{Phys. Rev. D} \textbf{\bibinfo{volume}{85}},
  \bibinfo{pages}{116002} (\bibinfo{year}{1998}),
  \eprint{arXiv:hep-ph/9809521}.

\bibitem[{\citenamefont{Bailey and
  Kosteleck{\'{y}}}(2006)}]{baileykostelecky2006}
\bibinfo{author}{\bibfnamefont{Q.~G.} \bibnamefont{Bailey}} \bibnamefont{and}
  \bibinfo{author}{\bibfnamefont{V.~A.} \bibnamefont{Kosteleck{\'{y}}}},
  \bibinfo{journal}{Phys. Rev. D} \textbf{\bibinfo{volume}{74}},
  \bibinfo{pages}{045001} (\bibinfo{year}{2006}),
  \eprint{arXiv:hep-ph/0407252}.

\bibitem[{\citenamefont{Kosteleck{\'{y}} and
  Tasson}(2009)}]{kosteleckytasson2009}
\bibinfo{author}{\bibfnamefont{V.~A.} \bibnamefont{Kosteleck{\'{y}}}}
  \bibnamefont{and} \bibinfo{author}{\bibfnamefont{J.~D.}
  \bibnamefont{Tasson}}, \bibinfo{journal}{Phys. Rev. Lett.}
  \textbf{\bibinfo{volume}{102}}, \bibinfo{pages}{010402}
  (\bibinfo{year}{2009}), \eprint{arXiv:0810.1459 [gr-qc]}.

\bibitem[{\citenamefont{Kosteleck{\'{y}} and
  Russell}(2011)}]{kosteleckyrussell2011}
\bibinfo{author}{\bibfnamefont{V.~A.} \bibnamefont{Kosteleck{\'{y}}}}
  \bibnamefont{and} \bibinfo{author}{\bibfnamefont{N.}~\bibnamefont{Russell}},
  \bibinfo{journal}{Rev. Mod. Phys.} \textbf{\bibinfo{volume}{83}},
  \bibinfo{pages}{11} (\bibinfo{year}{2011}), \eprint{arXiv:0801.0287
  [hep-ph]}.

\bibitem[{\citenamefont{Bear et~al.}(2000)\citenamefont{Bear, Stoner,
  Walsworth, Kosteleck{{\'{y}}}, and Lane}}]{bestwakola2000}
\bibinfo{author}{\bibfnamefont{D.}~\bibnamefont{Bear}},
  \bibinfo{author}{\bibfnamefont{R.~E.} \bibnamefont{Stoner}},
  \bibinfo{author}{\bibfnamefont{R.~L.} \bibnamefont{Walsworth}},
  \bibinfo{author}{\bibfnamefont{V.~A.} \bibnamefont{Kosteleck{{\'{y}}}}},
  \bibnamefont{and} \bibinfo{author}{\bibfnamefont{C.~D.} \bibnamefont{Lane}},
  \bibinfo{journal}{Phys. Rev. Lett.} \textbf{\bibinfo{volume}{85}},
  \bibinfo{pages}{5038} (\bibinfo{year}{2000}), \bibinfo{note}{{E}rratum,
  ibid., 89, 209902 (2002)}, \eprint{arXiv:physics/0007049 [physics.atom-ph]}.

\bibitem[{\citenamefont{Phillips et~al.}(2001)\citenamefont{Phillips, Humphrey,
  Mattison, Stoner, Vessot, and Walsworth}}]{phhumastvewa2001}
\bibinfo{author}{\bibfnamefont{D.~F.} \bibnamefont{Phillips}},
  \bibinfo{author}{\bibfnamefont{M.~A.} \bibnamefont{Humphrey}},
  \bibinfo{author}{\bibfnamefont{E.~M.} \bibnamefont{Mattison}},
  \bibinfo{author}{\bibfnamefont{R.~E.} \bibnamefont{Stoner}},
  \bibinfo{author}{\bibfnamefont{R.~F.~C.} \bibnamefont{Vessot}},
  \bibnamefont{and} \bibinfo{author}{\bibfnamefont{R.~L.}
  \bibnamefont{Walsworth}}, \bibinfo{journal}{Phys. Rev. D}
  \textbf{\bibinfo{volume}{63}}, \bibinfo{pages}{111101}
  (\bibinfo{year}{2001}), \eprint{arXiv:physics/0008230 [physics.atom-ph]}.

\bibitem[{\citenamefont{Humphrey et~al.}(2003)\citenamefont{Humphrey, Phillips,
  Mattison, Vessot, Stoner, and Walsworth}}]{huphmavest2003}
\bibinfo{author}{\bibfnamefont{M.~A.} \bibnamefont{Humphrey}},
  \bibinfo{author}{\bibfnamefont{D.~F.} \bibnamefont{Phillips}},
  \bibinfo{author}{\bibfnamefont{E.~M.} \bibnamefont{Mattison}},
  \bibinfo{author}{\bibfnamefont{R.~F.~C.} \bibnamefont{Vessot}},
  \bibinfo{author}{\bibfnamefont{R.~E.} \bibnamefont{Stoner}},
  \bibnamefont{and} \bibinfo{author}{\bibfnamefont{R.~L.}
  \bibnamefont{Walsworth}}, \bibinfo{journal}{Phys. Rev. A}
  \textbf{\bibinfo{volume}{68}}, \bibinfo{pages}{063807}
  (\bibinfo{year}{2003}), \eprint{arXiv:physics/0103068 [physics.atom-ph]}.

\bibitem[{\citenamefont{M{\"u}ller
  et~al.}(2003{\natexlab{a}})\citenamefont{M{\"u}ller, Braxmaier, Herrmann,
  Peters, and L{\"a}mmerzahl}}]{mubrhepela2003}
\bibinfo{author}{\bibfnamefont{H.}~\bibnamefont{M{\"u}ller}},
  \bibinfo{author}{\bibfnamefont{C.}~\bibnamefont{Braxmaier}},
  \bibinfo{author}{\bibfnamefont{S.}~\bibnamefont{Herrmann}},
  \bibinfo{author}{\bibfnamefont{A.}~\bibnamefont{Peters}}, \bibnamefont{and}
  \bibinfo{author}{\bibfnamefont{C.}~\bibnamefont{L{\"a}mmerzahl}},
  \bibinfo{journal}{Phys. Rev. D} \textbf{\bibinfo{volume}{67}},
  \bibinfo{pages}{056006} (\bibinfo{year}{2003}{\natexlab{a}}),
  \eprint{arXiv:hep-ph/0212289}.

\bibitem[{\citenamefont{M{\"u}ller
  et~al.}(2003{\natexlab{b}})\citenamefont{M{\"u}ller, Herrmann, Saenz, Peters,
  and L{\"a}mmerzahl}}]{muhesapela2003}
\bibinfo{author}{\bibfnamefont{H.}~\bibnamefont{M{\"u}ller}},
  \bibinfo{author}{\bibfnamefont{S.}~\bibnamefont{Herrmann}},
  \bibinfo{author}{\bibfnamefont{A.}~\bibnamefont{Saenz}},
  \bibinfo{author}{\bibfnamefont{A.}~\bibnamefont{Peters}}, \bibnamefont{and}
  \bibinfo{author}{\bibfnamefont{C.}~\bibnamefont{L{\"a}mmerzahl}},
  \bibinfo{journal}{Phys. Rev. D} \textbf{\bibinfo{volume}{68}},
  \bibinfo{pages}{116006} (\bibinfo{year}{2003}{\natexlab{b}}),
  \eprint{arXiv:hep-ph/0401016}.

\bibitem[{\citenamefont{M{\"u}ller}(2005)}]{muller2005}
\bibinfo{author}{\bibfnamefont{H.}~\bibnamefont{M{\"u}ller}},
  \bibinfo{journal}{Phys. Rev. D} \textbf{\bibinfo{volume}{71}},
  \bibinfo{pages}{045004} (\bibinfo{year}{2005}),
  \eprint{arXiv:hep-ph/0412385}.

\bibitem[{\citenamefont{Russell}(2005)}]{russell2005}
\bibinfo{author}{\bibfnamefont{N.}~\bibnamefont{Russell}},
  \bibinfo{journal}{Phys. Scripta} \textbf{\bibinfo{volume}{72}},
  \bibinfo{pages}{C38} (\bibinfo{year}{2005}), \eprint{arXiv:hep-ph/0501127}.

\bibitem[{\citenamefont{Casana et~al.}(2008)\citenamefont{Casana, {Ferreira
  Jr.}, and Santos}}]{casanaferreirasantos2008a}
\bibinfo{author}{\bibfnamefont{R.}~\bibnamefont{Casana}},
  \bibinfo{author}{\bibfnamefont{M.~M.} \bibnamefont{{Ferreira Jr.}}},
  \bibnamefont{and} \bibinfo{author}{\bibfnamefont{C.~E.~H.}
  \bibnamefont{Santos}}, \bibinfo{journal}{Phys. Rev. D}
  \textbf{\bibinfo{volume}{78}}, \bibinfo{pages}{025030}
  (\bibinfo{year}{2008}), \eprint{arXiv:0804.0431 [hep-th]}.

\bibitem[{\citenamefont{Casana et~al.}(2009)\citenamefont{Casana, {Ferreira
  Jr.}, Rodrigues, and Silva}}]{casanaferreirarodriguessilva2009}
\bibinfo{author}{\bibfnamefont{R.}~\bibnamefont{Casana}},
  \bibinfo{author}{\bibfnamefont{M.~M.} \bibnamefont{{Ferreira Jr.}}},
  \bibinfo{author}{\bibfnamefont{J.~S.} \bibnamefont{Rodrigues}},
  \bibnamefont{and} \bibinfo{author}{\bibfnamefont{M.~R.~O.}
  \bibnamefont{Silva}}, \bibinfo{journal}{Phys. Rev. D}
  \textbf{\bibinfo{volume}{80}}, \bibinfo{pages}{085026}
  (\bibinfo{year}{2009}), \eprint{arXiv:0907.1924 [hep-th]}.

\bibitem[{\citenamefont{Bakke et~al.}(2011)\citenamefont{Bakke, Belich, and
  Silva.}}]{bakkebelichsilva2011a}
\bibinfo{author}{\bibfnamefont{K.}~\bibnamefont{Bakke}},
  \bibinfo{author}{\bibfnamefont{H.}~\bibnamefont{Belich}}, \bibnamefont{and}
  \bibinfo{author}{\bibfnamefont{E.~O.} \bibnamefont{Silva.}},
  \bibinfo{journal}{J. Math. Phys.} \textbf{\bibinfo{volume}{52}},
  \bibinfo{pages}{063505} (\bibinfo{year}{2011}), \eprint{arXiv:1106.2324
  [hep-th]}.

\bibitem[{\citenamefont{Belich et~al.}(2011)\citenamefont{Belich, Silva,
  {Ferreira Jr.}, and Orlando}}]{besifeor2011}
\bibinfo{author}{\bibfnamefont{H.}~\bibnamefont{Belich}},
  \bibinfo{author}{\bibfnamefont{E.~O.} \bibnamefont{Silva}},
  \bibinfo{author}{\bibfnamefont{M.~M.} \bibnamefont{{Ferreira Jr.}}},
  \bibnamefont{and} \bibinfo{author}{\bibfnamefont{M.~T.~D.}
  \bibnamefont{Orlando}}, \bibinfo{journal}{Phys. Rev. D}
  \textbf{\bibinfo{volume}{83}}, \bibinfo{pages}{125025}
  (\bibinfo{year}{2011}), \eprint{arXiv:1106.0789 [hep-th]}.

\bibitem[{\citenamefont{Lima et~al.}(2013)\citenamefont{Lima, Belich, and
  Bakke}}]{limabelichbakke2013}
\bibinfo{author}{\bibfnamefont{A.~G.} \bibnamefont{Lima}},
  \bibinfo{author}{\bibfnamefont{H.}~\bibnamefont{Belich}}, \bibnamefont{and}
  \bibinfo{author}{\bibfnamefont{K.}~\bibnamefont{Bakke}},
  \bibinfo{journal}{Eur. Phys. J. Plus} \textbf{\bibinfo{volume}{128}},
  \bibinfo{pages}{154} (\bibinfo{year}{2013}).

\bibitem[{\citenamefont{Hernaski and Belich}(2014)}]{hernaskibelich2014}
\bibinfo{author}{\bibfnamefont{C.~A.} \bibnamefont{Hernaski}} \bibnamefont{and}
  \bibinfo{author}{\bibfnamefont{H.}~\bibnamefont{Belich}},
  \bibinfo{journal}{Phys. Rev. D} \textbf{\bibinfo{volume}{89}},
  \bibinfo{pages}{104027} (\bibinfo{year}{2014}), \eprint{arXiv:1409.5742
  [hep-th]}.

\bibitem[{\citenamefont{Berger and
  Kosteleck{\'y}}(2002)}]{bergerkostelecky2002}
\bibinfo{author}{\bibfnamefont{M.~S.} \bibnamefont{Berger}} \bibnamefont{and}
  \bibinfo{author}{\bibfnamefont{V.~A.} \bibnamefont{Kosteleck{\'y}}},
  \bibinfo{journal}{Phys. Rev. D} \textbf{\bibinfo{volume}{65}},
  \bibinfo{pages}{091701} (\bibinfo{year}{2002}),
  \eprint{arXiv:hep-th/0112243}.

\bibitem[{\citenamefont{Nibbelink and Pospelov}(2005)}]{nibbelinkpospelov2005}
\bibinfo{author}{\bibfnamefont{S.~G.} \bibnamefont{Nibbelink}}
  \bibnamefont{and} \bibinfo{author}{\bibfnamefont{M.}~\bibnamefont{Pospelov}},
  \bibinfo{journal}{Phys. Rev. Lett.} \textbf{\bibinfo{volume}{94}},
  \bibinfo{pages}{081601} (\bibinfo{year}{2005}),
  \eprint{arXiv:hep-ph/0404271}.

\bibitem[{\citenamefont{Katz and Shadmi}(2006)}]{katzshadmi2006}
\bibinfo{author}{\bibfnamefont{A.}~\bibnamefont{Katz}} \bibnamefont{and}
  \bibinfo{author}{\bibfnamefont{Y.}~\bibnamefont{Shadmi}},
  \bibinfo{journal}{Phys. Rev. D} \textbf{\bibinfo{volume}{74}},
  \bibinfo{pages}{115021} (\bibinfo{year}{2006}),
  \eprint{arXiv:hep-ph/0605210}.

\bibitem[{\citenamefont{Helay{\"{e}}l-{N}eto
  et~al.}(2010)\citenamefont{Helay{\"{e}}l-{N}eto, Belich, Dias, Leal, and
  Spalenza}}]{hnbedilesp2010}
\bibinfo{author}{\bibfnamefont{J.~A.} \bibnamefont{Helay{\"{e}}l-{N}eto}},
  \bibinfo{author}{\bibfnamefont{H.}~\bibnamefont{Belich}},
  \bibinfo{author}{\bibfnamefont{G.~S.} \bibnamefont{Dias}},
  \bibinfo{author}{\bibfnamefont{F.~J.~L.} \bibnamefont{Leal}},
  \bibnamefont{and} \bibinfo{author}{\bibfnamefont{W.}~\bibnamefont{Spalenza}},
  \bibinfo{journal}{Proc. Science} \textbf{\bibinfo{volume}{032}}
  (\bibinfo{year}{2010}).

\bibitem[{\citenamefont{Farias et~al.}(2012)\citenamefont{Farias, Lehum,
  Nascimento, and Petrov}}]{falenape2012}
\bibinfo{author}{\bibfnamefont{C.~F.} \bibnamefont{Farias}},
  \bibinfo{author}{\bibfnamefont{A.~C.} \bibnamefont{Lehum}},
  \bibinfo{author}{\bibfnamefont{J.~R.} \bibnamefont{Nascimento}},
  \bibnamefont{and} \bibinfo{author}{\bibfnamefont{A.~Y.}
  \bibnamefont{Petrov}}, \bibinfo{journal}{Phys. Rev. D}
  \textbf{\bibinfo{volume}{86}}, \bibinfo{pages}{065035}
  (\bibinfo{year}{2012}), \eprint{arXiv:1206.4508 [hep-th]}.

\bibitem[{\citenamefont{Redigolo}(2012)}]{redigolo2012}
\bibinfo{author}{\bibfnamefont{D.}~\bibnamefont{Redigolo}},
  \bibinfo{journal}{Phys. Rev. D} \textbf{\bibinfo{volume}{85}},
  \bibinfo{pages}{085009} (\bibinfo{year}{2012}), \eprint{arXiv:1106.2035
  [hep-th]}.

\bibitem[{\citenamefont{Gomes et~al.}(2013)\citenamefont{Gomes, Lehum,
  Nascimento, Petrov, and {d}a {S}ilva}}]{golenapeds2013}
\bibinfo{author}{\bibfnamefont{M.}~\bibnamefont{Gomes}},
  \bibinfo{author}{\bibfnamefont{A.~C.} \bibnamefont{Lehum}},
  \bibinfo{author}{\bibfnamefont{J.~R.} \bibnamefont{Nascimento}},
  \bibinfo{author}{\bibfnamefont{A.~Y.} \bibnamefont{Petrov}},
  \bibnamefont{and} \bibinfo{author}{\bibfnamefont{A.~J.} \bibnamefont{{d}a
  {S}ilva}}, \bibinfo{journal}{Phys. Rev. D} \textbf{\bibinfo{volume}{87}},
  \bibinfo{pages}{027701} (\bibinfo{year}{2013}), \eprint{arXiv:1210.6863
  [hep-th]}.

\bibitem[{\citenamefont{Belich et~al.}(2015)\citenamefont{Belich, Bernald,
  Gaete, Helay{\"e}l{-}Neto, and Leal}}]{bebegahnle2015}
\bibinfo{author}{\bibfnamefont{H.}~\bibnamefont{Belich}},
  \bibinfo{author}{\bibfnamefont{L.~D.} \bibnamefont{Bernald}},
  \bibinfo{author}{\bibfnamefont{P.}~\bibnamefont{Gaete}},
  \bibinfo{author}{\bibfnamefont{J.~A.} \bibnamefont{Helay{\"e}l{-}Neto}},
  \bibnamefont{and} \bibinfo{author}{\bibfnamefont{F.~J.~L.}
  \bibnamefont{Leal}}, \bibinfo{journal}{Eur. Phys. J. C}
  \textbf{\bibinfo{volume}{75}}, \bibinfo{pages}{291} (\bibinfo{year}{2015}),
  \eprint{arXiv:1502.06126 [hep-th]}.

\bibitem[{\citenamefont{Carroll et~al.}(1990)\citenamefont{Carroll, Field, and
  Jackiw}}]{cafija90}
\bibinfo{author}{\bibfnamefont{S.~M.} \bibnamefont{Carroll}},
  \bibinfo{author}{\bibfnamefont{G.~B.} \bibnamefont{Field}}, \bibnamefont{and}
  \bibinfo{author}{\bibfnamefont{R.}~\bibnamefont{Jackiw}},
  \bibinfo{journal}{Phys. Rev. D} \textbf{\bibinfo{volume}{41}},
  \bibinfo{pages}{1231} (\bibinfo{year}{1990}).

\bibitem[{\citenamefont{Dunne}(1998)}]{dunne1998}
\bibinfo{author}{\bibfnamefont{G.}~\bibnamefont{Dunne}},
  \emph{\bibinfo{title}{Aspects of the Chern-Simons theory in topological
  aspects of low dimensional systems}} (\bibinfo{publisher}{Springer},
  \bibinfo{address}{Berlin}, \bibinfo{year}{1998}).

\bibitem[{\citenamefont{Belich et~al.}(2013)\citenamefont{Belich, Bernald,
  Gaete, and Helay{\"e}l{-}Neto}}]{bebegahn2013}
\bibinfo{author}{\bibfnamefont{H.}~\bibnamefont{Belich}},
  \bibinfo{author}{\bibfnamefont{L.~D.} \bibnamefont{Bernald}},
  \bibinfo{author}{\bibfnamefont{P.}~\bibnamefont{Gaete}}, \bibnamefont{and}
  \bibinfo{author}{\bibfnamefont{J.~A.} \bibnamefont{Helay{\"e}l{-}Neto}},
  \bibinfo{journal}{Eur. Phys. J. C} \textbf{\bibinfo{volume}{73}},
  \bibinfo{pages}{2632} (\bibinfo{year}{2013}), \eprint{arXiv:1303.1108
  [hep-th]}.

\bibitem[{\citenamefont{Bonetti
  et~al.}(2016{\natexlab{a}})\citenamefont{Bonetti, {dos Santos Filho},
  {Helay\"{e}l-Neto}, and Spallicci}}]{bodshnsp2016}
\bibinfo{author}{\bibfnamefont{L.}~\bibnamefont{Bonetti}},
  \bibinfo{author}{\bibfnamefont{L.~R.} \bibnamefont{{dos Santos Filho}}},
  \bibinfo{author}{\bibfnamefont{J.~A.} \bibnamefont{{Helay\"{e}l-Neto}}},
  \bibnamefont{and} \bibinfo{author}{\bibfnamefont{A.~D. A.~M.}
  \bibnamefont{Spallicci}}, \emph{\bibinfo{title}{Photon sector analysis of
  {S}uper and {L}orentz symmetry breaking}}
  (\bibinfo{year}{2016}{\natexlab{a}}), \bibinfo{note}{in preparation}.

\bibitem[{\citenamefont{Adam and Klinkhamer}(2001)}]{adamklinkhamer2001}
\bibinfo{author}{\bibfnamefont{C.}~\bibnamefont{Adam}} \bibnamefont{and}
  \bibinfo{author}{\bibfnamefont{F.~R.} \bibnamefont{Klinkhamer}},
  \bibinfo{journal}{Nucl. Phys. B} \textbf{\bibinfo{volume}{607}},
  \bibinfo{pages}{247} (\bibinfo{year}{2001}), \eprint{arXiv:hep-ph/0306245}.

\bibitem[{\citenamefont{{Ba{{\^e}}ta Scarpelli}
  et~al.}(2003)\citenamefont{{Ba{{\^e}}ta Scarpelli}, Belich, Boldo, and
  {Helay{\"{e}}ël-Neto}}}]{bsbebohn2003}
\bibinfo{author}{\bibfnamefont{A.~P.} \bibnamefont{{Ba{{\^e}}ta Scarpelli}}},
  \bibinfo{author}{\bibfnamefont{H.}~\bibnamefont{Belich}},
  \bibinfo{author}{\bibfnamefont{J.~L.} \bibnamefont{Boldo}}, \bibnamefont{and}
  \bibinfo{author}{\bibfnamefont{J.~A.} \bibnamefont{{Helay{\"{e}}ël-Neto}}},
  \bibinfo{journal}{Phys. Rev. D} \textbf{\bibinfo{volume}{67}},
  \bibinfo{pages}{085021} (\bibinfo{year}{2003}),
  \eprint{arXiv:hep-th/0204232}.

\bibitem[{\citenamefont{Gomes and Malta}(2016)}]{gomesmalta2016}
\bibinfo{author}{\bibfnamefont{Y.~M.~P.} \bibnamefont{Gomes}} \bibnamefont{and}
  \bibinfo{author}{\bibfnamefont{P.~C.} \bibnamefont{Malta}},
  \emph{\bibinfo{title}{Lab-based limits on the {C}arroll-{F}ield-{J}ackiw
  {L}orentz violating electroynamics}} (\bibinfo{year}{2016}),
  \eprint{arXiv:1604.01102 [hep-ph]. To appear in Phys. Rev. D}.

\bibitem[{\citenamefont{Bentum et~al.}(2016)\citenamefont{Bentum, Bonetti, and
  Spallicci}}]{bebosp2016}
\bibinfo{author}{\bibfnamefont{M.~J.} \bibnamefont{Bentum}},
  \bibinfo{author}{\bibfnamefont{L.}~\bibnamefont{Bonetti}}, \bibnamefont{and}
  \bibinfo{author}{\bibfnamefont{A.~D. A.~M.} \bibnamefont{Spallicci}}
  (\bibinfo{year}{2016}), \bibinfo{note}{submitted; see
  https://www.utwente.nl/ctit/cat/Projects/OLFAR/}.

\bibitem[{\citenamefont{{de Broglie}}(1940)}]{db40}
\bibinfo{author}{\bibfnamefont{L.}~\bibnamefont{{de Broglie}}},
  \emph{\bibinfo{title}{La m\'echanique ondulatoire du photon, Une novelle
  th\'eorie de la lumi\'ere}} (\bibinfo{publisher}{Hermann},
  \bibinfo{address}{Paris}, \bibinfo{year}{1940}).

\bibitem[{\citenamefont{Williams et~al.}(1971)\citenamefont{Williams, Faller,
  and Hill}}]{wifahi71}
\bibinfo{author}{\bibfnamefont{E.~R.} \bibnamefont{Williams}},
  \bibinfo{author}{\bibfnamefont{J.~E.} \bibnamefont{Faller}},
  \bibnamefont{and} \bibinfo{author}{\bibfnamefont{H.~A.} \bibnamefont{Hill}},
  \bibinfo{journal}{Phys. Rev. Lett.} \textbf{\bibinfo{volume}{26}},
  \bibinfo{pages}{721} (\bibinfo{year}{1971}).

\bibitem[{\citenamefont{Ryutov}(1997)}]{ry97}
\bibinfo{author}{\bibfnamefont{D.~D.} \bibnamefont{Ryutov}},
  \bibinfo{journal}{Plasma Phys. Contr. Fus.} \textbf{\bibinfo{volume}{39}},
  \bibinfo{pages}{A73} (\bibinfo{year}{1997}).

\bibitem[{\citenamefont{Ryutov}(2007)}]{ry07}
\bibinfo{author}{\bibfnamefont{D.~D.} \bibnamefont{Ryutov}},
  \bibinfo{journal}{Plasma Phys. Contr. Fus.} \textbf{\bibinfo{volume}{49}},
  \bibinfo{pages}{B429} (\bibinfo{year}{2007}).

\bibitem[{\citenamefont{{Olive} and {the Particle Data
  Group}}(2014)}]{oliveetal2014}
\bibinfo{author}{\bibfnamefont{K.~A.} \bibnamefont{{Olive}}} \bibnamefont{and}
  \bibinfo{author}{\bibnamefont{{the Particle Data Group}}},
  \bibinfo{journal}{Chin. Phys. C} \textbf{\bibinfo{volume}{38}},
  \bibinfo{pages}{090001} (\bibinfo{year}{2014}).

\bibitem[{\citenamefont{Retin{\`o} et~al.}(2016)\citenamefont{Retin{\`o},
  Spallicci, and Vaivads}}]{retinospalliccivaivads2016}
\bibinfo{author}{\bibfnamefont{A.}~\bibnamefont{Retin{\`o}}},
  \bibinfo{author}{\bibfnamefont{A.~D. A.~M.} \bibnamefont{Spallicci}},
  \bibnamefont{and} \bibinfo{author}{\bibfnamefont{A.}~\bibnamefont{Vaivads}},
  \bibinfo{journal}{Astropart. Phys.} \textbf{\bibinfo{volume}{82}},
  \bibinfo{pages}{49} (\bibinfo{year}{2016}), \eprint{arXiv:1302.6168
  [hep-ph]}.

\bibitem[{\citenamefont{Goldhaber and Trimble}(2010)}]{goldhabertrimble1996}
\bibinfo{author}{\bibfnamefont{M.}~\bibnamefont{Goldhaber}} \bibnamefont{and}
  \bibinfo{author}{\bibfnamefont{V.}~\bibnamefont{Trimble}},
  \bibinfo{journal}{J. Astrophys. Astron.} \textbf{\bibinfo{volume}{17}},
  \bibinfo{pages}{17} (\bibinfo{year}{2010}).

\bibitem[{\citenamefont{Bonetti
  et~al.}(2016{\natexlab{b}})\citenamefont{Bonetti, Ellis, Mavromatos,
  Sakharov, {Sarkisian-Grinbaum}, and Spallicci}}]{boelmasasgsp2016}
\bibinfo{author}{\bibfnamefont{L.}~\bibnamefont{Bonetti}},
  \bibinfo{author}{\bibfnamefont{J.}~\bibnamefont{Ellis}},
  \bibinfo{author}{\bibfnamefont{N.~E.} \bibnamefont{Mavromatos}},
  \bibinfo{author}{\bibfnamefont{A.~S.} \bibnamefont{Sakharov}},
  \bibinfo{author}{\bibfnamefont{E.~K.} \bibnamefont{{Sarkisian-Grinbaum}}},
  \bibnamefont{and} \bibinfo{author}{\bibfnamefont{A.~D. A.~M.}
  \bibnamefont{Spallicci}}, \bibinfo{journal}{Phys. Lett. B}
  \textbf{\bibinfo{volume}{757}}, \bibinfo{pages}{548}
  (\bibinfo{year}{2016}{\natexlab{b}}), \eprint{arXiv: 1602.09135
  [astro-ph.HE]}.

\bibitem[{\citenamefont{Wu et~al.}(2016)\citenamefont{Wu, Zhang, Gao, Wei, Zou,
  Lei, Zhang, Dai, and M{\'{e}}sz{\'{a}}ros}}]{wuetal2016b}
\bibinfo{author}{\bibfnamefont{X.}~\bibnamefont{Wu}},
  \bibinfo{author}{\bibfnamefont{S.}~\bibnamefont{Zhang}},
  \bibinfo{author}{\bibfnamefont{H.}~\bibnamefont{Gao}},
  \bibinfo{author}{\bibfnamefont{J.}~\bibnamefont{Wei}},
  \bibinfo{author}{\bibfnamefont{Y.}~\bibnamefont{Zou}},
  \bibinfo{author}{\bibfnamefont{W.}~\bibnamefont{Lei}},
  \bibinfo{author}{\bibfnamefont{B.}~\bibnamefont{Zhang}},
  \bibinfo{author}{\bibfnamefont{Z.}~\bibnamefont{Dai}}, \bibnamefont{and}
  \bibinfo{author}{\bibfnamefont{P.}~\bibnamefont{M{\'{e}}sz{\'{a}}ros}},
  \bibinfo{journal}{Astophys. J. Lett.} \textbf{\bibinfo{volume}{822}},
  \bibinfo{pages}{L15} (\bibinfo{year}{2016}), \eprint{arXiv:1602.07835
  [astro-ph.HE]}.

\end{thebibliography}

\end{document}